# Custodial and non-custodial wallets

## Tony Seymour and Geoff Goodell

Non-custodial wallets are a type of cryptocurrency wallet wherein the owner has full control over the private keys and is solely responsible for managing and securing the digital assets that it contains [1]. Unlike custodial wallets, which are managed by third parties, such as exchanges, non-custodial wallets ensure that funds are controlled exclusively by the end user.

The difference between custodial and non-custodial wallets contrasts with the distinction between assets that are held on an owner's device and assets that are held by a third party. The distinction is related to the difference between control and possession:

• **control:** "transferable property of a relationship between an actor and an asset wherein the actor and no other actor has the means to specify legitimate changes to the asset" [2]
• **possession:** "transferable property of a relationship between an actor and an asset wherein the actor and no other actor can effect changes to the asset" [2]

|  | held by owner | held by third party |
|---|---|---|
| **third party in control** | wallet with a hardware root of trust that enforces rules on behalf of an authority or issuer | online services that store digital assets and conduct transactions on behalf of the owner |
| **owner in control** | hardware or software wallet accountable solely to the owner | online services that store digital assets but for which the owner is responsible for their management and any transactions |

Figure 1.  Control versus possession.  [2]

"Control differs from possession in that possession determines who can access an asset, whereas control determines who can use an asset. It is possible to have possession without control, or control without possession" [2] (see Figure 1).

Custodial wallets typically involve services managed by a third-party exchange that holds and manages digital assets on behalf of the user and provides a software wallet to facilitate the use of these services. However, some exchanges and services might offer hardware custodial wallet solutions, for example, to enforce rules about how the assets are used.  Irrespective of whether the wallet is implemented in software or hardware, the question of whether it is custodial or non-custodial is a matter of who is in control of the assets.  If the assets are within the security envelope of a service provider, then the service provider is in control.

**Key points about custodial wallets:**

1. **Control of Private Keys**: The service provider or authority controls the private keys, not the user. This means the provider is responsible for the security and management of the user's funds.
2. **Recovery and Support**: Depending upon whether the service provider is in possession, users may be able to rely on the service provider for account recovery in case of lost assets. This contrasts with non-custodial wallets, where losing private keys usually means losing access to funds.
3. **Security Risks**: While reputable custodial services employ strong security measures, they are also targets for hackers. Users must trust that the provider can safeguard their assets against potential breaches.
4. **Limited Control**: Users may have limited control over their funds. For example, the provider may impose withdrawal limits, fees, or freeze assets under certain conditions.
5. **Regulation and Compliance**: Custodial wallet providers are often regulated and may be legally required to comply with Know Your Customer (KYC) and Anti-Money Laundering (AML) rules that require monitoring the transactions made by their wallets. This could affect privacy.
6. **Custody Fees**: Some custodial wallets charge fees for managing funds or for specific transactions, making them potentially more expensive than non-custodial options.

Some examples of custodial wallet providers:

1. **Coinbase**: Primarily a software wallet, but they offer a hardware wallet integration through Ledger.
2. **Binance**: Similar to Coinbase, Binance offers custodial services and can integrate with hardware wallets like Ledger and Trezor.
3. **Kraken**: Another exchange that provides custodial wallet services and supports hardware wallet integrations

**Key points about non-custodial wallets:**

1. **Full Control**: The owner possesses the private keys that confer exclusive control over the assets. Depending upon the design of the system, the owner may also possess the assets themselves.
2. **Security**: Since no third party has access to the owner's keys, the security of the owner's assets depends on how well the owner manages the owner's keys.
3. **Privacy**: Non-custodial wallets can often offer greater privacy since they do not necessarily share personally identifiable information (PII) when transacting.
4. **Compatibility**: Non-custodial wallets can interact with decentralized finance (DeFi) protocols and decentralized applications without relying upon a third party service provider.

Some examples of non-custodial wallets :

- **MetaMask**: A software wallet that is used to interact with dApps and DeFi.
- **Ledger**: A hardware wallet known for its security.
- **Trezor**: Another hardware wallet with strong security features.

These tools allow users to manage the private keys directly, providing convenience but requiring users trust the security of their implementation.

**Risks of using custodial wallets**

Using custodial wallets comes with several risks, primarily because the owner is entrusting a third party with the control of the private keys. Here are some of the main risks:

1. **Security Breaches**: If the custodial service is hacked, the owner's assets could be stolen [3]. This has happened in the past with several high-profile exchange hacks.
2. **Bankruptcy or Insolvency**: If the company managing a custodial wallet goes bankrupt, the owner might lose access to the owner's assets [3].
3. **Lack of Control**: Because the owner does not have full control over the private keys, the owner must rely on the service provider to manage and secure the assets.
4. **Regulatory Risks**: Custodial services are subject to regulatory scrutiny and could be forced to seize assets in its possession or freeze assets under its control if required by law.
5. **Service Limitations**: Some custodial wallets may only support specific types of cryptocurrencies or transactions, limiting the owner's flexibility.

Since custodial wallets implicitly rely upon accounts, procedures for enhancing the security of a custodial wallet and mitigating the risks associated with accounts are crucial to protect the digital assets. Some steps are shown below:

1. **Enable Two-Factor Authentication (2FA)**: Combining multiple factors in an authentication process adds an extra layer of security by requiring a second form of verification, such as a code generated by or sent to a device or a mobile phone.
2. **Use Strong, Unique Passwords**: Where passwords are used, owners can create high-entropy passwords that contain a series of unrelated words or include a mix of letters, numbers, and symbols. Owners can avoid using the same password across multiple platforms.
3. **Be Wary of Phishing Attacks**: Owners can adopt the practice of double-checking URLs and avoiding clicking on suspicious links or downloading unknown software.
4. **Regularly Monitor Your Account**: Owners can monitor their account activity and, if facilitated by the service provider, the owner can set up alerts for any unusual transactions.
5. **Use Non-Custodial Wallets for Security of Large Holdings**: Owners can choose to store a significant portion of their assets in non-custodial wallets, even if they choose to use custodial services for convenience when transacting.
6. **Use Non-Custodial Wallets for Privacy of Transactions**: If possible, the owner can choose to transfer assets to a non-custodial wallet for conducting transactions.
7. **Keep Software Updated**: The owner can establish procedures for updating wallet software, firmware (if applicable), and any related applications to ensure that they are are always up to date, to protect against vulnerabilities.

## Risks associated with using non-custodial wallets:

1. **Loss of Private Keys**:
   - The most significant risk is that if an owner loses access to the private keys (or recovery phrase) or the assets themselves (if applicable), then the owner loses access to the assets permanently. Unlike custodial wallets, there is no recovery option for forgetting or losing credentials, since there is no third party to which to authenticate.
2. **Human Error**:
   - Managing private keys can be complicated, especially for beginners. Mistakes such as sending funds to the wrong address, improperly backing up keys, or exposing private keys to malicious actors can result in loss of assets.
3. **Security Vulnerabilities**:
   - While non-custodial wallets can be highly secure, they are only as secure as the user's practices. A poorly secured device, weak password, or lack of proper backup can expose the wallet to theft or hacking.
4. **Phishing Attacks and Scams**:

- Since users handle private keys directly, owners might be more vulnerable to phishing attacks and scams designed to trick them into revealing sensitive information or downloading malware that compromises the wallet.

5. **Device Loss or Damage**:
   - If an owner stores a non-custodial wallet on a mobile device or uses a hardware wallet, and the owner then loses or damages the device without properly backing up the assets, the assets can become irretrievable.

6. **Lack of Customer Support**:
   - Non-custodial wallets typically offer no or very limited customer support. If an owner encounters technical issues or have questions, the only recourse might be to rely on community forums or self-help resources.

7. **Software Vulnerabilities**:
   - Non-custodial wallets might depend on software or applications, which in turn might have bugs or vulnerabilities. If the wallet software is compromised or not updated frequently, it could expose users to potential attacks.

8. **Responsibility for Security**:
   - Unlike custodial wallets where a service provider is responsible for security, in non-custodial wallets, the user is solely responsible for protecting private keys and wallet access. This necessitates an understanding of security best practices, such as enabling two-factor authentication (2FA) and ensuring proper backups.

9. **No Recourse for Errors**:
   - Transactions on a ledger are irreversible. If an owner makes a mistake, such as sending cryptocurrency to the wrong address or accidentally relinquishing control of assets in the wallet, then there is no third party to help recover the assets.

10. **Potential Hardware Vulnerabilities**:
    - For users of hardware wallets (a type of non-custodial wallet), the risk of hardware failure, tampering, or malfunction could lead to issues accessing funds if backups are not maintained properly.

## Summary of custodial and non-custodial wallets

### 1. Control of Private Keys

- **Custodial Wallets**: The third-party provider (such as an exchange or wallet service) controls and manages the private keys on behalf of the user. The user essentially trusts the provider with the security and accessibility of the assets.
- **Non-Custodial Wallets**: The user has full control over the private keys (and possibly the assets themselves). This means only the owner is in control of their assets, making the owner responsible for security and management.

### 2. Security

- **Custodial Wallets**: Security depends on the service provider. If the provider is hacked, compromised, or experiences a security breach, the user's funds could be at risk. However, many large custodial services offer insurance and security measures, such as two-factor authentication.
- **Non-Custodial Wallets**: Security is in the hands of the user. If private keys are stored securely (e.g., using hardware wallets or strong encryption), they can be very secure. However, losing access to the private key (or recovery phrase, if applicable) can result in the permanent loss of the assets.

### 3. Ease of Use

- **Custodial Wallets**: The third party handles the technical aspects, such as private key management, backups, and security. Users can recover their accounts if they forget their passwords.
- **Non-Custodial Wallets**: Managing a non-custodial wallet can require more technical knowledge. In particular, owners must securely manage their private keys, and there is no recovery option if private keys are lost.

## 4. Privacy

- **Custodial Wallets**: Because they rely upon accounts, custodial wallet services typically require users to complete Know Your Customer (KYC) processes, meaning users must provide personal information such as their name, address, and identification. This results in less privacy.
- **Non-Custodial Wallets**: Users retain full privacy over their wallet. Since they control their private keys, they are not required to submit personal information (depending on how and where the wallet is used) and can avoid both revealing PII when transacting and revealing information about their transaction history to service providers.

## 5. Ownership of Assets

- **Custodial Wallets**: Since the provider holds the private keys, the user does not have direct control over their funds. In extreme cases, the provider may limit withdrawals, impose fees, or freeze accounts, for example, to enforce regulatory compliance or address security concerns.
- **Non-Custodial Wallets**: Users have complete ownership of their funds and control transactions without any intermediary. There are no limits imposed by third parties, ensuring full autonomy over assets.

## 6. Backup and Recovery

- **Custodial Wallets**: The provider manages backup and recovery via an account. If the user forgets their password or login credentials, the provider typically offers a recovery process.
- **Non-Custodial Wallets**: The user is solely responsible for backing up their private keys or recovery phrases. If these are lost, there is no way to recover the wallet.

## 7. Security Risks

- **Custodial Wallets**: Custodial wallets are at risk of being hacked, compromised, or affected by internal issues (for example, insolvency of the service provider). High-profile hacks and thefts have occurred at exchanges and custodial services.
- **Non-Custodial Wallets**: Security risks primarily come from user error, such as mishandling private keys, falling victim to phishing, or using insecure devices.

## 9. Regulation and Compliance

- **Custodial Wallets**: Because they involve account provision, custodial wallets are often subject to government regulations and compliance rules (such as KYC and AML). This may lead to the seizure or freezing of assets or the imposition of withdrawal or transaction limits.
- **Non-Custodial Wallets**: Because there is no regulated service provider, non-custodial wallets are generally not subject to regulatory oversight, allowing users more flexibility and autonomy over their assets.